\documentclass[aps,prb,twocolumn]{revtex4}
\usepackage{graphicx}
\begin{document}
\title{Ab initio study of charge order in Fe$_{3}$O$_{4}$ 
}

\author{Z. Szotek$^{1}$, W.M. Temmerman$^{1}$, A. Svane$^{2}$, 
L. Petit$^{2}$, G.M. Stocks$^{3}$, and H. Winter$^{4}$}
\affiliation{$^{1}$ Daresbury Laboratory, Daresbury, Warrington, WA4 4AD, Cheshire, U.K. \\
$^{2}$ Institute of Physics and Astronomy, University of Aarhus, DK-8000 Aarhus C, Denmark\\
$^{3}$ Metals and Ceramics Division, Oak Ridge National Laboratory, Oak Ridge, TN, USA \\
$^{4}$ INFP, Forschungszentrum Karlsruhe GmbH, Postfach 3640, D-76021 Karlsruhe, Germany}

\date{\today}

\begin{abstract}
We present a self-interaction corrected local spin density (SIC-LSD) study of
the electronic structure and possible charge order of magnetite, Fe$_{3}$O$_{4}$.
The issue of charge order in magnetite is explored in both cubic and orthorhombic 
structures, the latter being an approximation to the true, low temperature, 
monoclinic structure. We find that the Verwey charge ordered phase is not the 
groundstate solution for this compound either in cubic or orthorhombic structure.
We conclude that the structural distortions, more than localization/delocalization
correlations, are responsible for the charge disproportionation in the low temperature
phase.
\end{abstract}

\maketitle

\section{Introduction}
Based upon its high magnetoresistive properties, magnetite, Fe$_{3}$O$_{4}$, is 
a system of interest for such technological applications as computer memory, 
magnetic recording, and other "spintronic" devices. It has been extensively
studied for the past 60 years\cite{Walz}, but despite such an extended effort the issue
of a possible charge order below the so-called Verwey transition temperature,
T$_{V}$=122 K, has not been understood yet\cite{Verwey}. Above the Verwey transition
temperature, magnetite is thought to be a half-metal with a highest known 
T$_{c}$ of 860 K. At the transition
temperature, the conductivity of the system decreases abruptly by two orders
of magnitude, as a small gap opens up in the density of states spectrum.\cite{Park} This first 
order metal-insulator transition \cite{Imada} is associated with a lattice distortion 
from cubic to monoclinic, which however has not been fully resolved.\cite{Izumi,Zuo,Paolo} 
The transition has been viewed as an order-disorder transition in relation to the 
arrangement of cations on the octahedral sites of the inverse spinel structure whose 
formal chemical formula can be written as 
Fe$_{A}$$^{3+}$ [ Fe$^{2+}$, Fe$^{3+}$ ]$_{B}$O$_{4}$$^{2-}$, where $A$ and $B$
refer to the interstitial, i.e., tetrahedral and octahedral sites, respectively. 
This simple ionic picture implies that magnetite is a mixed-valent compound. 
The $A$ sublattice is occupied only by Fe$^{3+}$ cations in a ferromagnetic arrangement,
while the $B$ sublattice is occupied by both Fe$^{2+}$ (B1-site) and Fe$^{3+}$ (B2-site)
cations, also in a ferromagnetic manner which, however, is antiferromagnetically aligned 
to the ferromagnetic order on the $A$ sublattice. The oxygen atoms occupy the 
face-centered-cubic ({\it fcc}) 
close-packed sites. The Fe$^{2+}$ cation can be viewed as Fe$^{3+}$ plus an "extra" 
electron, hopping freely between the octahedral sites 
above the Verwey transition, while below the transition these "extra" electrons become
localized or quasi-localized and some kind of long range order sets in 
on the $B$ sublattice. Verwey, who 
argued that below the transition temperature, T$_{V}$, the Fe$^{3+}$ and Fe$^{2+}$ 
cations order in alternate (001) planes, looked upon this transition as an electron 
localization-delocalization transition. The M\"{o}ssbauer spectra clearly demonstrate 
existence of two types of localized ions on the octahedral sites.\cite{Moss} 
The low temperature structure is a rhombohedral distortion of the cubic spinel
arrangement to the first approximation, \cite{rook} but this splits in the ratio 3:1 
for the occurance of the Fe$^{2+}$ and Fe$^{3+}$ cations on the corner-sharing tetrahedra
that are formed by the octahedral sites. This is not consistent with the Anderson condition 
of minimum electrostatic repulsion, implying the ratio of 2:2. \cite{anderson} A recent
combined high resolution X-ray and neutron powder diffraction study \cite{Paolo} confirmed
a small charge disproportionation of about 0.2 electron on the octahedral sites, and 
a violation of the Anderson condition. The current experimental situation suggests that
the long range order present below the Verwey transition is much more complicated than
the charge order first proposed by Verwey. Some experiments, 
claim that there exists no charge order on the octahedral sites.\cite{garcia1,garcia2}

In this paper we present an application of the self-interaction corrected local
spin density (SIC-LSD) approximation to Fe$_{3}$O$_{4}$. 
The SIC-LSD method allows the ab-initio realization of valencies through
total energy calculations for configurations of Fe$^{2+}$ and Fe$^{3+}$ cations.
The method includes all bandstructure effects of the delocalized Fe and O electrons,
including their hybridization, as well as the electronic structure of the localized
Fe electrons. As such this methodology has been able to describe the valencies 
of the rare earths successfully.\cite{Strange}
Here, we concentrate
on the total energies for different $d$ electron configurations on the octahedral
sites, defining Fe valencies, and try to establish whether the simple charge order
postulated by Verwey is the ground state solution. Apart from the Verwey
order we also consider other obvious arrangements of Fe valencies on the octahedral sites.
There have been other first-principles studies of magnetite, \cite{Sashy,Anisimov,Antonov}
however, this is the first time that different Fe valencies and charge order scenarios
have been studied by an ab initio method. In Refs. \onlinecite{Anisimov} and \onlinecite{Antonov},
the LDA+U method has been used to study the Verwey insulating phase albeit in the cubic
structure only, and neither the total energies nor other charge order phases have been considered. 
In Ref. \onlinecite{Antonov}, in addition, the optical properties of magnetite have been studied.

The remainder of the paper is organized as follows. In the next section, we briefly present the 
theoretical background of the SIC-LSD approach. In section III, the technical and computational
details regarding the application to magnetite are elaborated upon. In section IV, we 
present total energy and density of states calculations. We also study the charge order
and different Fe valencies on the octahedral sites, and try to establish the ground state 
configuration. The paper is concluded in section V.

\section{Theory}

The basis of the SIC-LSD formalism is a self-interaction free total energy functional, 
\( E^{SIC} \), obtained by subtracting from the LSD total energy functional,
\( E^{LSD} \), a spurious self-interaction of each occupied electron state 
\( \psi _{\alpha } \)\cite{PZ81}, namely 
\begin{equation}
\label{eq1}
E^{SIC}=E^{LSD}-\sum _{\alpha }^{occ.}\delta _{\alpha }^{SIC}.
\end{equation}
 Here \( \alpha  \) numbers the occupied states and the self-interaction correction
for the state \( \alpha  \) is 
\begin{equation}
\delta _{\alpha }^{SIC}=U[n_{\alpha }]+E_{xc}^{LSD}[\bar{n}_{\alpha }],
\end{equation}
with \( U[n_{\alpha }] \) being the Hartree energy and \( E_{xc}^{LSD}[\bar{n}_{\alpha }] \)
the LSD exchange-correlation energy for the corresponding charge density \( n_{\alpha } \)
and spin density \( \bar{n}_{\alpha } \). 

The SIC-LSD approach can be viewed as an extension of LSD
in the sense that the self-interaction correction is only finite for spatially
localized states, while for Bloch-like single-particle states \( E^{SIC} \)
is equal to \( E^{LSD} \). Thus, the LSD minimum is also a local minimum of
\( E^{SIC} \). A question now arises, whether there exist other competitive
minima, corresponding to a finite number of localized states, which could benefit
from the self-interaction term without loosing too much 
of the energy associated with band formation.
This is often the case for rather well localized electrons like the 3 $d$
electrons in transition metal oxides or the 4\( f \) electrons in rare earth
compounds. It follows from minimization of Eq. (\ref{eq1}) that within the SIC-LSD 
approach
such localized electrons move in a different potential than the delocalized
valence electrons which respond to the effective LSD potential. For example,
in the case of magnetite, five (Fe$^{3+}$) or six (Fe$^{2+}$)
Fe $d$ electrons move in the SIC potential, while all other electrons feel 
only the effective LSD potential. Thus, by including
an explicit energy contribution for an electron to localize, the ab-initio SIC-LSD
describes both localized and delocalized electrons on an equal footing, leading
to a greatly improved description of static Coulomb correlation effects over
the LSD approximation. 

In order to make the connection between valency and localization more explicit
it is useful to define the nominal valency as
\[
N_{val}=Z-N_{core}-N_{SIC},
\]
where $Z$ is the atomic number (26 for Fe), $N_{core}$ is the number of core 
(and semi-core) electrons (18 for Fe), and $N_{SIC}$ is the number of localized, 
i.e., self-interaction corrected, states (either five or six, respectively for 
Fe$^{3+}$ and Fe$^{2+}$). Thus, in this formulation the valency is equal to the 
integer number of electrons available for band formation. To find the valency we 
assume various atomic configurations, consisting of different numbers of localized 
states, and minimize the SIC-LSD energy functional of Eq. (\ref{eq1}) with respect 
to the number of localized electrons. The SIC-LSD formalism is governed by the 
energetics due to the fact that for each orbital the SIC differentiates between 
the energy gain due to hybridization of the orbital with the valence bands and the 
energy gain upon localization of the orbital. Whichever wins determines if the
orbital is part of the valence band or not and in this manner
also leads to the evaluation of the valency of elements involved.
The SIC depends on the choice of orbitals and its value
can differ substantially as a result of this. Therefore, one
has to be guided by the energetics in defining the most optimally
localized orbitals to determine the absolute energy minimum of
the SIC-LSD energy functional. The advantage of the SIC-LSD formalism is that 
for such systems as transition metal oxides or rare earth compounds the lowest 
energy solution will describe the situation where some single-electron states 
may not be Bloch-like. For magnetite, these would be the Fe 3$d$ states, but 
not the O 2$p$ states, as trying to localize the latter is energetically unfavourable.

In the present work the SIC-LSD approach, has been implemented\cite{TSSW98} 
within the linear muffin-tin-orbital (LMTO) atomic sphere approximation (ASA) band 
structure method,\cite{oka75} in the 
tight-binding representation.\cite{AJ84} The electron wave functions are expanded in
terms of the screened muffin-tin orbitals, and the minimization of \( E^{SIC} \)
becomes non-linear in the expansion coefficients.  

\section{Calculational details}

As mentioned earlier, magnetite crystallizes in the {\it fcc} inverse spinel structure 
with two formula units (14 atoms) in the primitive unit cell, and the space group is 
{\it Fd$\overline{3}$m}. In the actual calculations, 
an additional 16 empty spheres have been used in order to obtain better
space filling and to minimize the ASA overlap. Concerning the linear muffin-tin 
basis functions, we have used $4s$, $4p$, and $3d$ partial waves on all Fe atoms, and treated
them as low-waves. \cite{Lambrecht} Including also $4f-$basis functions on the Fe-sites, 
treated as intermediate waves, is of no substantial importance for the final results. 
On the oxygens only $2s$ and $2p$ partial waves have been treated as low-waves, while the 
$3d-$waves have been treated as intermediate. On the empty spheres the $1s-$waves have been 
considered as low- and the $2p$'s as intermediate-waves. The empty spheres are  
necessary to obtain half-metallic solution with the LSD, whose self-consistent charge 
densities have been used to start SIC-LSD calculations. 

Although, below the Verwey 
transition, the cubic structure undergoes a monoclinic distortion, we have first explored 
the charge ordered phases on the basis of the cubic structure and then extended the study 
to the distorted structure. Since the true monoclinic structure has
not been fully resolved, our calculations for the low temperature phase have been
performed for two orthorhombic structures, given in Refs. \onlinecite{Izumi} and 
\onlinecite{Paolo}, as closest approximations to the monoclinic structure. In the
latter work, the refined atom positions were obtained using a combination of  high resolution 
X-ray and neutron powder diffraction. The respective space groups used are 
Pmc2$_1$ and P2/c, each with 56 atoms in the unit cell. However, for reliable
calculations empty spheres have been included, increasing the total number of sites to
104 in the first case, and 120 for the refined structure of Wright et al. \cite{Paolo}

To make the SIC-LSD calculations feasible, for the orthorhombic set-ups we had to
restrict to minimum basis sets, in particular $4s, 4p, 3d-$low waves have been used on 
every Fe-atom, $2s, 2p-$low waves and intermediate $3d-$waves on all oxygen atoms and only
$1s-$low waves and intermediate $2p-$waves on the empty spheres. Also, the number of k-points 
in the irreducible Brillouin zone (BZ) had to be reduced to 18 k-points.

To study charge order, we have implemented three obvious scenarios. First, we have considered 
the simple Verwey charge order, where it is assumed that the tetrahedral sites are occupied 
by Fe$^{3+}$ ions, while on all the octahedral sites Fe alternates between Fe$^{3+}$ (five $d$ 
electrons move in the SIC-LSD potential) and Fe$^{2+}$ (six $d$ electrons move in the SIC-LSD 
potential) in the successive (001) planes. Subsequently, we refer to this as Verwey charge order 
scenario. In the second scenario, we assume that all tetrahedral and octahedral sites are 
trivalent, namely Fe$^{3+}$. Finally, in the third scenario we assume all the tetrahedral 
sites to be Fe$^{3+}$, and all the octahedral sites to be Fe$^{2+}$. In what follows, we shall 
refer to these calculations as scenarios 1-3, respectively. 

Note that for a divalent Fe atom one needs to consider six $d$ electrons moving in the 
SIC potential. Five of the electrons fill up the full manifold of one spin component, 
whilst the sixth electron has the freedom to populate either of the five states of the 
other spin component (three t$_{2g}$ and two e$_{g}$ states in the cubic symmetry). 
As a result five separate, self-consistent total energy calculations have to be performed,
for all different structures and scenarios involving divalent Fe's, to find the lowest 
energy solution. In what follows, only the results of the lowest energy solutions are 
presented and discussed.

\section{Results and discussion}

\subsection{Densities of states}

In Fig. 1 we compare for the cubic structure the LSD total and species-decomposed densities 
of states (DOS) with the SIC-LSD counterparts where the Verwey charge order has been
implemented. In this figure both spin components are displayed. In Fig. 2, 
we compare separately the LSD spin-decomposed DOS for the tetrahedral and octahedral
sites with the respective DOS from the SIC-LSD calculations, where the octahedral sites
are additionally split into Fe$^{3+}$ and Fe$^{2+}$. 
The first thing to note about Fig. 1 is that the LSD
calculations find half-metallic DOS, while a small band gap opens up in both spin-channels
of the SIC-LSD densities of states, as a result of the Verwey charge order implemented on
the octahedral sites. The gap is about 0.35 eV, which is more than two
times the experimental gap found in photoemission for magnetite below the Verwey transition 
temperature, but in good agreement with the LDA+U calculation of Anisimov et al.,
\cite{Anisimov} although the SIC-LSD gap has the $p$-$d$ charge transfer character, while
the LDA+U gap occurs between $d$ states of the octahedral Fe ions. Note that
due to opening of the SIC-LSD gap, the unoccupied Fe 3$d$ states have been shifted up in energy, 
away from the Fermi level, while the occupied Fe 3$d$ states appear well below the valence 
band, which is composed mostly of oxygen 2$p$ with a small admixture of Fe $d$ states. In the LSD 
plots, the spin-up Fe 3$d$ states occur at the Fermi level, while there is a well defined gap in 
the other spin-component. In addition, the predominantly oxygen-like valence band is much
more structured in the case of LSD, due to strong hybridization with Fe 3$d$ states, which
appear at the top of the valence band. Looking at all different types of Fe sites in both
the calculations (see Fig. 2), it is clear that the Fe-A sites in the SIC-LSD
calculations do not contribute to DOS at the Fermi level, so there is a well defined gap
in both spin channels for these sites. The A-sites in the LSD calculation exhibit a very
small DOS in the spin-up component, and a gap in the other spin channel. The octahedral, 
B-sites, in the LSD calculations
exhibit a gap in one spin component and in the other spin-component well hybridized
states can be observed at the Fermi level, giving rise to the half-metallic ground state.
Among the octahedral sites in the SIC-LSD calculations, the B-sites occupied by Fe$^{3+}$
ions show similar DOS to the LSD calculations for the octahedral ions, except for the
gap appearing in the SIC-LSD case in both spin-channels, and the localized spin-down
Fe$^{3+}$ states occuring well below the valence bands. Regarding the unoccupied Fe 3$d$
states, the spin-up states for the Fe$^{2+}$ sites lie higher in energy than the 
corresponding states for the Fe$^{3+}$ sites. The small peak observed just below the
valence band is due to the sixth localized 3$d$ state, while the remaining five localized
$d-$states are visible at about 8.0 eV in the other spin component.

In Fig. 3 we show the total and partial Fe- and O-densities of states for the second
SIC-LSD scenario, namely for the case where all Fe-sites, tetrahedral and octahedral
alike, are 3+. The DOS are half-metallic, with a gap in one spin component and 
hybridized Fe-O states occuring at the Fermi level in the other spin component. The
DOS of localized Fe 3$d$ states appear well below the valence band at about 13.0
eV. Note substantial differences of these DOS with respect to the LSD result of Fig. 1.
The reason being, that in the LSD all electrons move in the effective LSD potential,
while in the 3+ scenario, the Fe 3$d$ electrons respond to the SIC potential.
Due to this localizing potential the total weight of the occupied 3$d$ states is 
shifted well below the valence band, while in the LSD picture they appear at the
top of the valence band, strongly hybridizing with the O 2$p$ states.
From the A and B-Fe site decomposed spin-polarized DOS in Fig. 4 we can see 
that the tetrahedral sites do not contribute at the Fermi level. The half-metallic 
behaviour is entirely due to the hybridization of the unoccupied 3$d$ states of the 
octahedral Fe sites with the oxygen 2$p$ states. 

Finally, the LSD and SIC-LSD densities of states for the orthorhombic structures 
(not displayed in this work) show
similar features to their counterparts in the cubic structure, when the respective
tetrahedral and octahedral ions are compared. Specifically, a small energy gap 
of about $0.1 eV$ and insulating solutions are observed in the charge ordered state  
for both orthorhombic symmetries. Note that in both orthorhombic structures that were 
studied there are six crystallographically different B-sites which are subsequently 
referred to as B1, ..., B6. \cite{Izumi}
Nevertheless, we still have only two types of Fe ions, divalent and trivalent.
Concerning the tetrahedral (A) sites, there are four different crystallographic positions
in the structure with Pmc2$_{1}$ symmetry, but only two A-type positions in the
structure with P2/c symmetry. 

Detailed comparison of the densities of states for all different crystallographic 
structures that we have implemented here, reveals that the second
scenario in the orthorhombic structure with the Pmc2$_{1}$ symmetry leads to
metallic rather than half-metallic DOS. This result is caused by the fact that the Fermi 
level moves slightly into the Fe-A 3$d$ peak, which is unoccupied in the 
cubic structure. However, for the refined orthorhombic structure of Ref. \cite{Paolo}, 
with P2/c symmetry, we again recover a half-metallic solution for the all trivalent 
configurations.
As in the cubic structure, the Fe-A 3$d$ peak is fully unoccupied. This
appears to indicate a great sensitivity of the half-metallic solution to atom positions 
in the low temperature structure.

\subsection{Total energies and charge order}

In this subsection we concentrate on the total energies, spin magnetic moments, and
charge disproportionation on octahedral sites, obtained for the scenarios studied here. 
Tables I and II summarize for the cubic and orthorhombic structures, respectively, the spin 
magnetic moments on Fe-sites and energy differences with respect to the ground state 
configuration of the most favourable scenario. 

Comparing total energies for 
the LSD and three different SIC-LSD charge order scenarios in either cubic (Table I) or 
orthorhombic (Table II) structure, one can see that the second scenario, with all Fe's 
in the trivalent configuration, is the lowest energy solution. It is followed by the Verwey charge
order scenario, then the third scenario, and finally the LSD, which is the most unfavourable result. 
The significant energy difference of 113 mRy per formula unit (Table I), found between the
scenario SIC(1) and scenario SIC(2), renders the simple Verwey ordering highly unlikely
as an explanation for the properties in the low temperature phase of magnetite.
The last two columns of Table I, $\Delta$E$_{BF}$ and $\Delta$E$_{SIC}$, representing respectively
the loss in band formation/hybridization energy and gain in localization energy, relative to
the respective values of the ground state configuration, provide further
physical understanding of this finding. Whilst the localization energy favours more localized
states, and hence the formation of Fe$^{2+}$ on the B2 octahedral sites, the loss of band formation 
energy by far outweighs the gain in localization energy. So, there is a strong competition between 
localization and band formation/hybridization energies, and in the simple Verwey scenario the loss 
in band formation energy, due to localization of one additional electron on the B2 octahedral sites,
is three times larger than the gain in localization energy. This balance could possibly be
reversed by increasing the nearest neighbour Fe (B2-type)-O bond length, which could perhaps be 
realized in a very distorted structure, although it is not the case for the orthorhombic structures
studied here. The expectation is that substantially increased Fe (B2-type)-O bond length would
lead to an increase in the localization energy of all six $d$ electrons of B2-Fe (with $2+$ valency),
and a reduction of their hybridization, thus establishing the Verwey phase as the lowest energy solution. 
However, one should keep in mind that we refer to nominal 
valencies, and therefore find that already Fe$^{3+}$ contains more than five $d$-electrons (Table III). 
All in all, Fe$_{3}$O$_{4}$ turns out to have mostly intermediate valencies.

Inspecting in detail the other
numbers of Table I, where the results for cubic structure are summarized, one can see that 
because these systems are either insulating or half-metallic their total magnetic moments are 
integer. The corresponding total spin magnetic moments for LSD, scenario SIC(1)
and scenario SIC(2) are equal to 4.0 $\mu_{B}$ per formula unit, while for the third scenario 
the total spin moment is 2.0 $\mu_{B}$ per formula unit. The latter value results from the
fact that for divalent ions, one additional $d$ electron becomes localized. The 
type-resolved octahedral sites spin magnetic moments are substantially larger for the 
trivalent ions than for the divalent ions in all the SIC-LSD calculations. The spin
magnetic moments of the tetrahedral Fe's are of similar magnitude but opposite sign
to the trivalent octahedral ions. As there are twice as many octahedral sites as 
tetrahedral ones, the total spin magnetic moment of magnetite in all the different 
scenarios is mostly due to the octahedral sites. The SIC-LSD spin
magnetic moments of the oxygens are of the order of 0.02-0.30 $\mu_{B}$, depending on the
scenario, while the spin magnetic moments on the empty spheres are still two orders of 
magnitude smaller. It is interesting that for the scenario with all octahedral ions in
divalent configuration all oxygen moments are oppositely polarized to the octahedral
Fe moments and are of the order of 0.2-0.3 $\mu_{B}$. This of course is reflected in
the much smaller total spin moment of this scenario. For the first and second scenarios
the O moments are smaller, between 0.04 and 0.17 $\mu_{B}$, and polarized parallel to 
the octahedral Fe ions. The oxygen moments calculated within LSD are of the
order of 0.1 $\mu_{B}$ and alligned anti-parallel to the octahedral Fe spin magnetic
moments.

\begin{table}
\caption{Total and type-decomposed spin magnetic moments (in Bohr magnetons 
$\mu_{B}$) for LSD and three different SIC-LSD scenarios for the cubic structure. 
Concerning types, only different Fe types are listed. Note that in the LSD,
SIC(2) and SIC(3)
calculations there is only one type of octahedral sites (B1$\equiv$B2$\equiv$B). 
Also given are 
the total energy differences (in mRy/formula unit) with respect to the ground 
state configuration. These energy differences are further decomposed into the energy
loss in band formation/hybridization ($\Delta$E$_{BF}$) and the energy gain due to
localization ($\Delta$E$_{SIC}$), both relatively to the respective energies of the
ground state configuration, which is the all Fe$^{3+}$ scenario.}
\begin{tabular}{cccccccc}
Scenario & M$_{total}$ & M$_{Fe_{A}}$ & M$_{Fe_{B1}}$ & M$_{Fe_{B2}}$ & $\Delta$E & $\Delta$E$_{BF}$ & $\Delta$E$_{SIC}$ \\
\hline
LSD     & 4.00 & 3.40 & -3.57 & -3.57 & 894 & -118 & 1012 \\
SIC (1) & 4.00 & 4.00 & -3.57 & -4.08 & 113 & 168 & -55 \\
SIC (2) & 4.00 & 4.02 & -3.90 & -3.90 & 0   & 0 & 0 \\
SIC (3) & 2.00 & 4.01 & -3.47 & -3.47 & 374 & 475 & -101 \\
\end{tabular}
\label{table1}
\end{table}

\begin{table}
\caption{Total and type-decomposed spin magnetic moments (in Bohr magnetons 
$\mu_{B}$) for LSD and three different SIC-LSD scenarios for the orthorhombic
structures. Concerning types, only different Fe types are listed. In the LSD
column for the P2/c  symmetry, 4.00(-) means that the magnetic moment is
a tiny fraction less, but within better than 0.5\% equal to 4.00 $\mu_{B}$. Also 
given are the total energy differences (in mRy/formula unit) with respect to the 
ground state configuration. 
}
\begin{tabular}{ccccc}
 & LSD & SIC (1) & SIC (2) & SIC (3) \\
\hline
  &  & Pmc2$_{1}$ symmetry  &  &   \\
\hline
M$_{total}$   & 4.00  & 4.00  & 4.02 & 2.00  \\
\hline
M$_{Fe_{A1}}$ & 3.39  & 4.13  &  4.14 & 4.14 \\
M$_{Fe_{A2}}$ & 3.32  & 4.13  &  4.13 & 4.15 \\
M$_{Fe_{A3}}$ & 3.30  & 4.14  &  4.16 & 4.17 \\
M$_{Fe_{A4}}$ & 3.34  & 4.15  &  4.16 & 4.16 \\
M$_{Fe_{B1}}$ & -3.40 & -3.58 & -3.88 & -3.50 \\
M$_{Fe_{B2}}$ & -3.46 & -4.14 & -3.93 & -3.51 \\
M$_{Fe_{B3}}$ & -3.39 & -4.11 & -3.90 & -3.54 \\
M$_{Fe_{B4}}$ & -3.43 & -3.58 & -3.98 & -3.43 \\
M$_{Fe_{B5}}$ & -3.41 & -4.13 & -3.95 & -3.46 \\
M$_{Fe_{B6}}$ & -3.37 & -4.11 & -3.94 & -3.48 \\
\hline
$\Delta$E & 966 & 105  & 0 & 327 \\
\hline
  &  & P2/c  symmetry &  &   \\
\hline
M$_{total}$ & 4.00(-) & 4.00 & 4.00 & 2.00  \\
\hline
M$_{Fe_{A1}}$ &  3.48 &  4.08 & 4.09 &  4.08 \\
M$_{Fe_{A2}}$ &  3.46 &  4.06 & 4.08 &  4.07 \\
M$_{Fe_{B1}}$ & -3.53 & -3.57 & 3.83 & -3.50 \\
M$_{Fe_{B2}}$ & -3.51 & -3.56 & 3.83 & -3.51 \\
M$_{Fe_{B3}}$ & -3.67 & -3.56 & 3.97 & -3.45 \\
M$_{Fe_{B4}}$ & -3.39 & -3.55 & 3.89 & -3.46 \\
M$_{Fe_{B5}}$ & -3.40 & -4.01 & 3.84 & -3.47 \\
M$_{Fe_{B6}}$ & -3.57 & -4.05 & 3.87 & -3.50 \\
\hline
$\Delta$E & 841 & 108  & 0  & 337  \\
\end{tabular}
\label{table2}
\end{table}

Turning to Table II (orthorhombic structure), one can see similar behaviour to that of Table I
(cubic structure). Except for the LSD,
the respective energy differences for different scenarios are slightly smaller, but otherwise
comparable to those for the cubic structure. Regarding the type of solution, metallic or
otherwise, again the scenario SIC(1) is insulating, for both orthorhombic structures, but
the second (all trivalent) scenario for Pmc2$_{1}$ symmetry and LSD for the P2/c symmetry
are not 100\% spin polarized, as reflected by their respective total spin magnetic
moments. Most importantly, however, for the refined atom positions, as reflected by the
structure with P2/c symmetry, \cite{Paolo} the all trivalent scenario is half-metallic.
Also, for the LSD solution in this structure we find only a minute deviation from 100\%
spin polarization with the DOS at the Fermi level of about 0.24 states per formula unit
and the total spin magnetic moment of almost 4.00 $\mu_{B}$ (with the accuracy better
than 0.5\%). Compared to the other orthorhombic
structure \cite{Izumi,Zuo} it appears that the refined atom positions are crucial for the
half-metallic solution in the second scenario. Regarding the type-resolved spin magnetic
moments, again we see that the trivalent tetrahedral and octahedral sites give rise to
comparable values, with substantially smaller moments of the divalent Fe octahedral sites.
Also, the oxygen spin moments are similar to those of the cubic structure.

\begin{table}
\caption{Listed are the numbers of spin-up and down valence electrons on the respective
Fe-atom types for the LSD (two types only) and three different SIC-LSD calculations
for the cubic structure.}
\begin{tabular}{ccccccc}
Scenario & ${Fe_{A}}^{\uparrow}$ & ${Fe_{A}}^{\downarrow}$ & ${Fe_{B1}}^{\uparrow}$ & ${Fe_{B1}}^{\downarrow}$ & ${Fe_{B2}}^{\uparrow}$ & ${Fe_{B2}}^{\downarrow}$  \\
\hline
LSD     & 5.06 & 1.67 & 2.15 & 5.71 & 2.15 & 5.71  \\
SIC (1) & 5.34 & 1.34 & 2.24 & 5.81 & 1.82 & 5.90  \\
SIC (2) & 5.35 & 1.33 & 1.97 & 5.87 & 1.97 & 5.87  \\
SIC (3) & 5.36 & 1.35 & 2.26 & 5.73 & 2.26 & 5.73  \\
\end{tabular}
\label{table3}
\end{table}

\begin{table}
\caption{Listed are the numbers of spin-up and down valence electrons on the respective 
Fe-atom types for the LSD (two types only) and three different SIC-LSD calculations for the 
orthorhombic structures.}
\begin{tabular}{ccccc}
 & LSD & SIC (1) & SIC (2) & SIC (3) \\
\hline
  &  & Pmc2$_{1}$ symmetry  &  &   \\
\hline
 ${Fe_{A1}}^{\uparrow}$   & 5.41 & 5.73 & 5.74 & 5.75 \\
 ${Fe_{A1}}^{\downarrow}$ & 2.02 & 1.60 & 1.60 & 1.60 \\
 ${Fe_{A2}}^{\uparrow}$   & 5.35 & 5.71 & 5.71 & 5.73 \\
 ${Fe_{A2}}^{\downarrow}$ & 2.03 & 1.58 & 1.58 & 1.58 \\
 ${Fe_{A3}}^{\uparrow}$   & 5.33 & 5.71 & 5.72 & 5.74 \\
 ${Fe_{A3}}^{\downarrow}$ & 2.03 & 1.57 & 1.56 & 1.57 \\
 ${Fe_{A4}}^{\uparrow}$   & 5.28 & 5.64 & 5.65 & 5.66 \\
 ${Fe_{A4}}^{\downarrow}$ & 1.94 & 1.49 & 1.49 & 1.50 \\
 ${Fe_{B1}}^{\uparrow}$   & 2.28 & 2.29 & 2.04 & 2.29 \\
 ${Fe_{B1}}^{\downarrow}$ & 5.68 & 5.87 & 5.92 & 5.79 \\
 ${Fe_{B2}}^{\uparrow}$   & 2.21 & 1.80 & 1.97 & 2.26 \\
 ${Fe_{B2}}^{\downarrow}$ & 5.67 & 5.94 & 5.90 & 5.77 \\
 ${Fe_{B3}}^{\uparrow}$   & 2.19 & 1.77 & 1.95 & 2.20 \\
 ${Fe_{B3}}^{\downarrow}$ & 5.58 & 5.88 & 5.85 & 5.74 \\
 ${Fe_{B4}}^{\uparrow}$   & 2.20 & 2.24 & 1.92 & 2.28 \\
 ${Fe_{B4}}^{\downarrow}$ & 5.63 & 5.82 & 5.90 & 5.71 \\
 ${Fe_{B5}}^{\uparrow}$   & 2.19 & 1.76 & 1.91 & 2.24 \\
 ${Fe_{B5}}^{\downarrow}$ & 5.60 & 5.89 & 5.86 & 5.70 \\
 ${Fe_{B6}}^{\uparrow}$   & 2.22 & 1.79 & 1.93 & 2.24 \\
 ${Fe_{B6}}^{\downarrow}$ & 5.59 & 5.90 & 5.87 & 5.72 \\
\hline
  &  & P2/c  symmetry &  &   \\
\hline
 ${Fe_{A1}}^{\uparrow}$   & 5.05 & 5.32 & 5.33 & 5.34 \\
 ${Fe_{A1}}^{\downarrow}$ & 1.57 & 1.24 & 1.24 & 1.26 \\
 ${Fe_{A2}}^{\uparrow}$   & 5.04 & 5.31 & 5.32 & 5.33 \\
 ${Fe_{A2}}^{\downarrow}$ & 1.58 & 1.25 & 1.24 & 1.26 \\
 ${Fe_{B1}}^{\uparrow}$   & 2.01 & 2.08 & 1.86 & 2.09 \\
 ${Fe_{B1}}^{\downarrow}$ & 5.54 & 5.65 & 5.69 & 5.59 \\
 ${Fe_{B2}}^{\uparrow}$   & 2.04 & 2.09 & 1.88 & 2.10 \\
 ${Fe_{B2}}^{\downarrow}$ & 5.55 & 5.65 & 5.71 & 5.61 \\
 ${Fe_{B3}}^{\uparrow}$   & 1.88 & 2.04 & 1.72 & 2.07 \\
 ${Fe_{B3}}^{\downarrow}$ & 5.55 & 5.60 & 5.69 & 5.52 \\
 ${Fe_{B4}}^{\uparrow}$   & 1.92 & 1.95 & 1.67 & 1.97 \\
 ${Fe_{B4}}^{\downarrow}$ & 5.31 & 5.50 & 5.56 & 5.43 \\
 ${Fe_{B5}}^{\uparrow}$   & 1.93 & 1.57 & 1.72 & 1.97 \\
 ${Fe_{B5}}^{\downarrow}$ & 5.33 & 5.58 & 5.56 & 5.44 \\
 ${Fe_{B6}}^{\uparrow}$   & 1.95 & 1.65 & 1.79 & 2.05 \\
 ${Fe_{B6}}^{\downarrow}$ & 5.52 & 5.70 & 5.66 & 5.55 \\
\end{tabular}
\label{table4}
\end{table}

The spin resolved charges for various Fe atoms for all scenarios and structures 
are given in Tables III and IV. The first thing to note is that except for the LSD scenario 
the electron charges of both spin components of the tetrahedral Fe's do not differ too much 
among various scenarios. The situation is different for the octahedral sites. Especially, 
in the Verwey charge ordered scenario the different octahedral ions show similar numbers 
in one spin component, but differ by about 0.42 electron in the other spin component
of the cubic structure, which is due to localization of the sixth $d$ electron. So, while 
in the ionic picture the valency of the two octahedral sites (B1 and B2)
differs by one, in terms of the charge disproportionation they are only 0.32 electron
different. The situation is similar for the orthorhombic structures. For the structure
with the Pmc2$_{1}$ symmetry the B1 and B4 sites are divalent, while B2, B3, B5 and B6 
sites are trivalent. In the other orthorhombic P2/c structure,
the sites B1 to B4 are divalent, while B5 and B6 sites are trivalent. Again, the 
charge of spin down components of all the octahedral sites is of the order of 5.8-5.9 
electron, the spin up charges of the divalent sites are about 2.2-2.3 electron, while 
for the trivalent sites they are of the order of 1.8 electron for the Pmc2$_{1}$ symmetry.
For the P2/c symmetry the spin down charges of the octahedral sites fall between 5.5 to
5.7 electron, while the spin down charges are of the order of 2.0 and 1.6-1.7 electron
for the divalent and trivalent octahedral ions respectively. Consequently, the total
charge disproportionation between the divalent and trivalent sites is up to 0.5 and 0.6
electron, respectively for the Pmc2$_{1}$ and P2/c symmetries. 
There is a spread in charges
over the octahedral sites of the orthorhombic structures
due to the reduced symmetry of these Fe sites.
In the orthorhombic structure we also see charge disproportionation in the other three
scenarios, namely LSD, scenario SIC(2) and SIC(3) - this of course does not occur in the cubic phase.
It is important to note that for the ground state configuration of all trivalent Fe's we find 
charge disproportionations of $0.2$ and $0.36$ electron for the
Pmc2$_{1}$ and P2/c symmetries respectively.
These calculated charge disproportionations are in better agreement with the recent
experimental study by Wright et al.\cite{Paolo} than the ones of the Verwey charge ordered state.

\section{Conclusions}

In summary, in the present application of the first-principles SIC-LSD method to 
magnetite we have studied three
different arrangements of Fe's on the octahedral sites, namely the simple Verwey
order, the case where, like the tetrahedral sites, all the octahedral sites 
have been occupied by Fe$^{3+}$, and finally the scenario with all the octahedral
sites occupied by the Fe$^{2+}$ ions. For all these scenarios the tetrahedral
sites have been occupied only by Fe$^{3+}$. Our total energy calculations, both for
the cubic and orthorhombic structures, have shown that the second scenario, with
all interstitial sites occupied by Fe$^{3+}$, is the ground state solution,
followed by the case of the Verwey charge order. In the latter case, we have
calculated magnetite to be insulating with a gap of about 0.35 eV ($\sim$0.1 eV in
the orthorhombic phase), while the
Fe$^{3+}$ scenario, depending on structure, has given rise to half-metallic
or metallic (though still with high spin polarization) state. The insulating state for
the Verwey scenario indicates that in the SIC-LSD description the O 2$p$ bands 
are filled, as would be the case for the ionic model. However, no ionic charge 
distribution is obtained as substantial Fe and O bonding and hybridization still
occur. Localizing one less Fe $d$ electron (second scenario) results in disappearance 
of the insulating state and emergence of the half-metallic state, which is a non-trivial
result, not predicted by the ionic model.

The ground state scenario of all trivalent Fe's gives a charge disproportionation which
is approximately half of what is found in the Verwey charge ordered state
in the orthorhombic structure.
The values of the ground state scenario are in line with the
recent experiments. \cite{Paolo,garcia1,garcia2} The charge disproportionation in the ground
state is similar to the one obtained from the LSD which would indicate that these 
charge disproportionations are not of an electronic origin but are determined by the structure.
Namely, the inequivalent Fe positions give rise to different charge distributions, whilst the
nominal Fe valency remains unaltered.

Obviously we have managed to study only the three simplest charge ordering scenarios, 
and within these restrictions our results indicate the energetically unfavourable character 
of Fe$^{2+}$. We also
find the same ground state configuration of all trivalent Fe in both the
cubic structure and the distorted orthorhombic structures.
The calculations indicate no minority spin electron localization in the distorted structure, only 
a marginal decrease, in this structure, with respect to the cubic one, 
of 5\% or so, in the energy difference between the ground state and the charge ordered one.
This implies that a simple charge ordered state as described by scenario 1 (SIC(1)) is not
realized below the Verwey temperature.

Our study has shown that the SIC-LSD is capable to 
study any kind of static charge order, although realization of more complicated 
charge arrangements, which might be of relevance to future studies,
 may be limited by memory and CPU of present computers.
Given these limitations that we only studied a simple Verwey charge ordered
state, we have concluded that the charge disproportionation below
the Verwey temperature is structural in origin and all Fe ions occur in a trivalent
state. This does not exclude the possibility that Fe$_{3}$O$_{4}$ below the Verwey 
temperature is described by a much more complicated charge ordered state.

\section*{Acknowledgements}
We thank Dr P.G. Radaelli for providing atom positions of the orthorhombic 
structure with P2/c symmetry, Professor Victor Antonov for useful communication,
 and Professor P.H. Dederichs for valuable discussions.
This work is partially funded by the Training and Mobility Network on 'Electronic
Structure Calculation of Materials Properties and Processes for Industry and Basic
Sciences' (contract:FMRX-CT98-0178) and the Research Training Network 'Psi-k f-electron'
(contract:HPRN-CT-2002-00295).
This work is partially sponsored by the Division of Materials Sciences and
Engineering, Office of Basic Energy Sciences, U.S. Department of Energy, under
Contract DE-AC05-00OR227725. The computational work reported here was
performed at the Center for Computational Sciences (CCS) at ORNL and at
NERSC at LBNL.\\


\begin{figure*}
\begin{tabular}{cc}
\includegraphics[scale=.37,angle=-90]{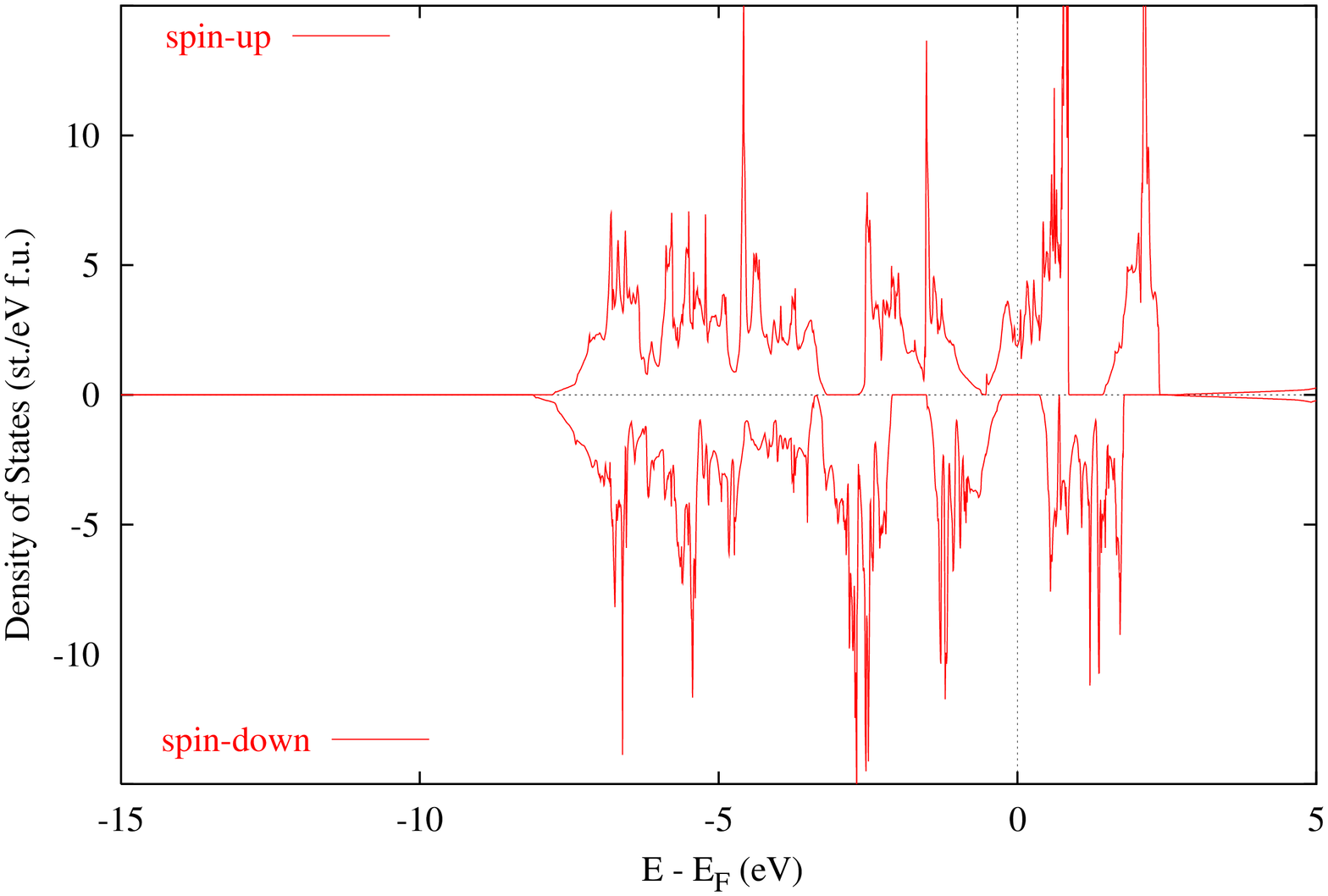}&
\includegraphics[scale=.37,angle=-90]{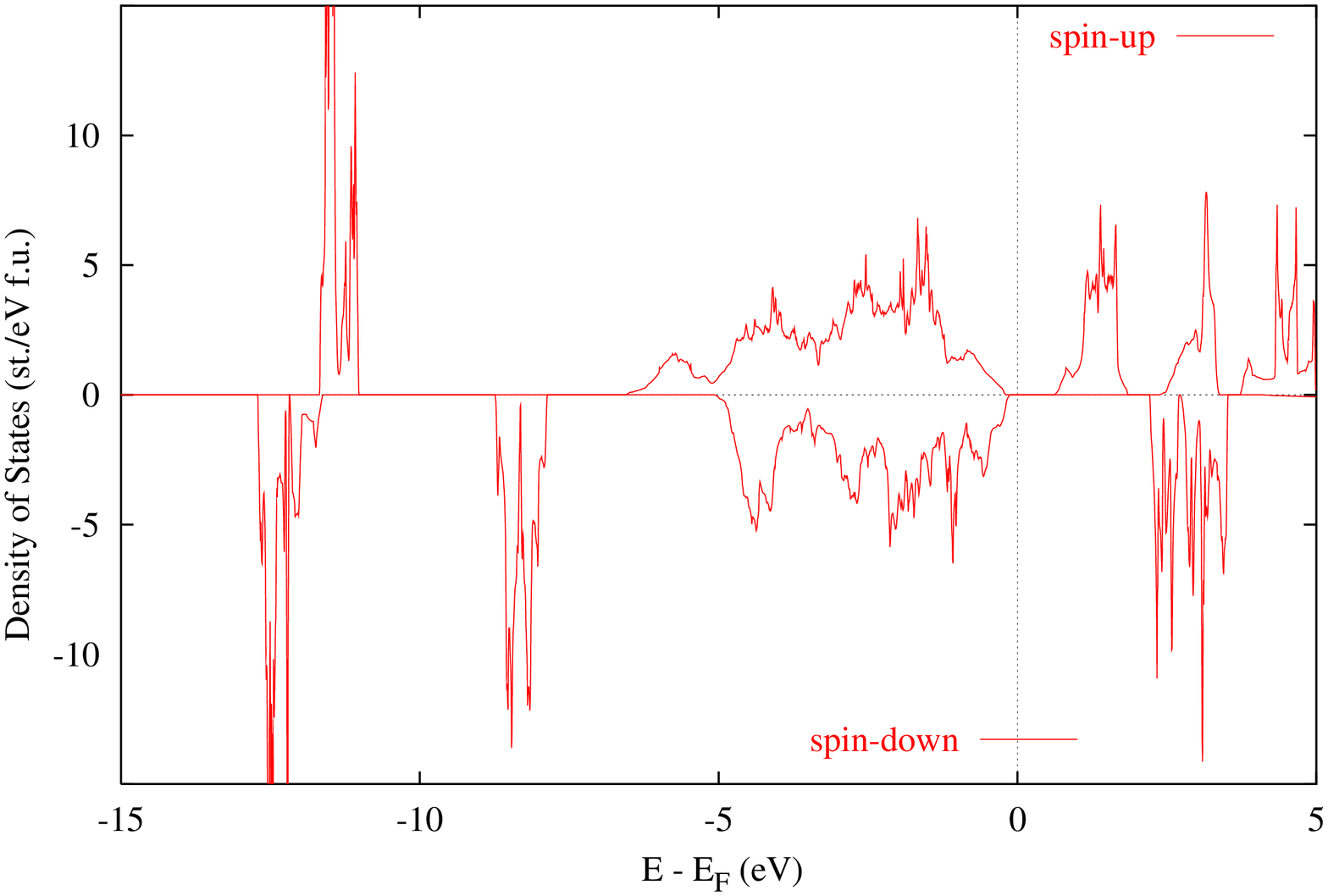}\\
\includegraphics[scale=.37,angle=-90]{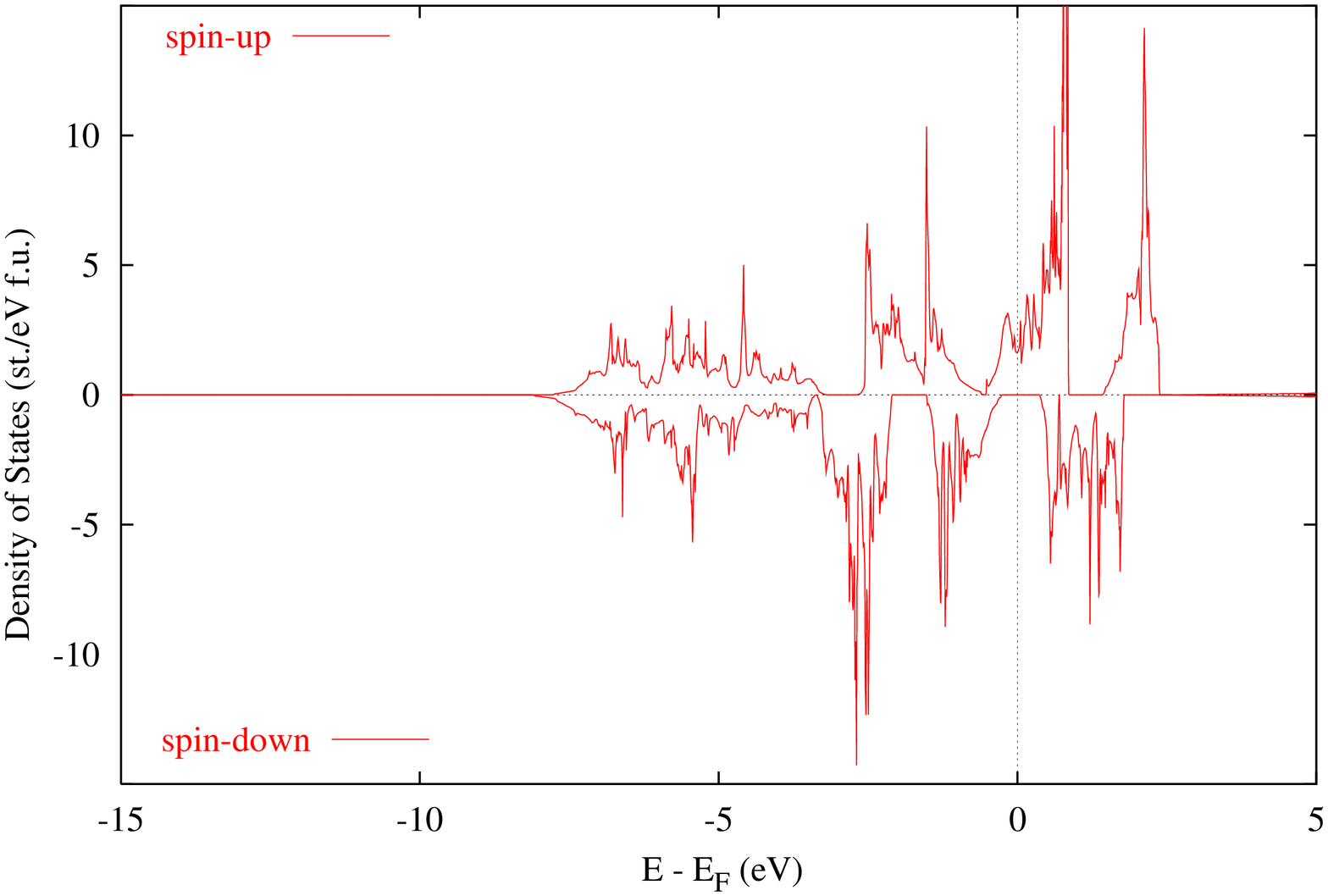}&
\includegraphics[scale=.37,angle=-90]{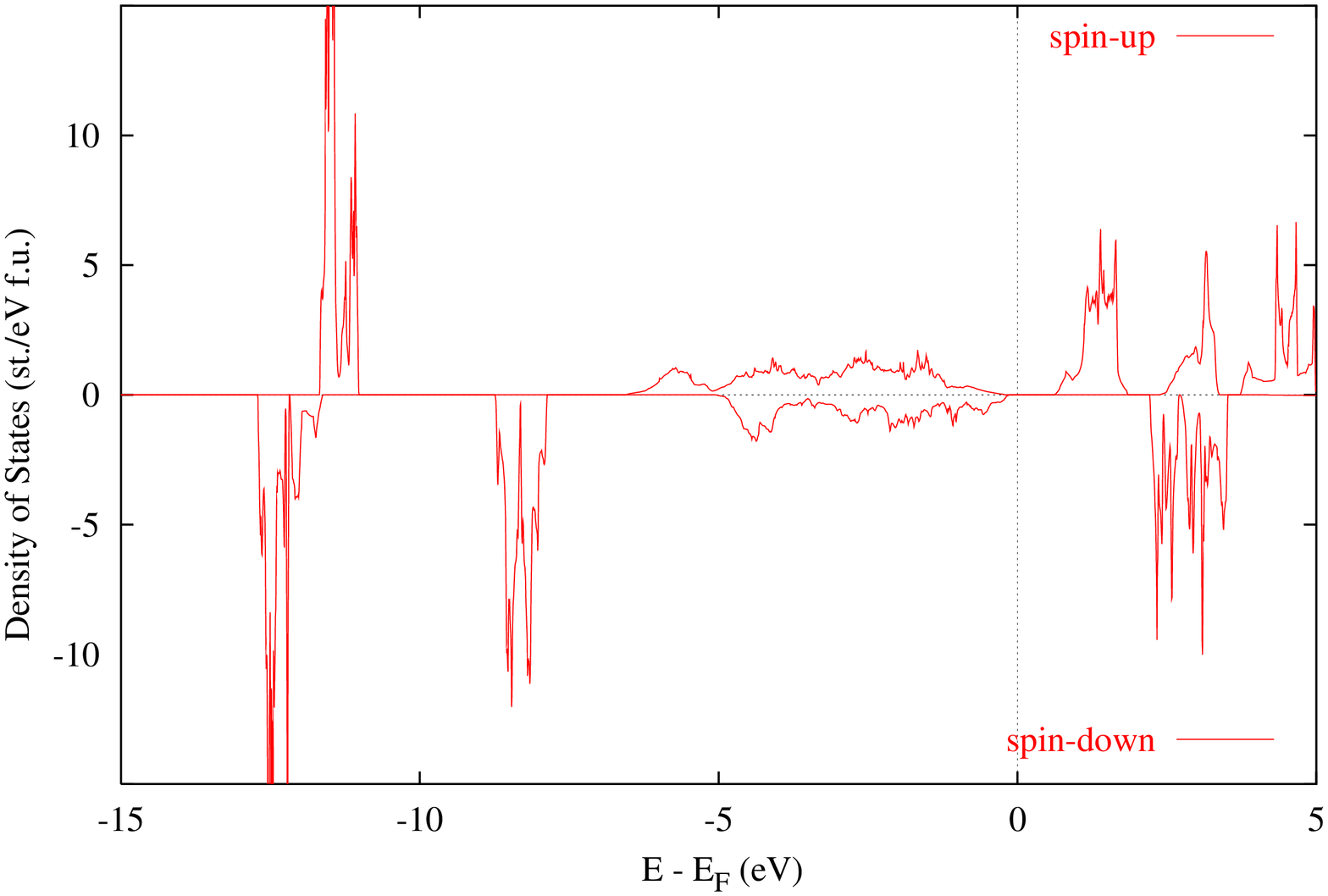}\\
\includegraphics[scale=.37,angle=-90]{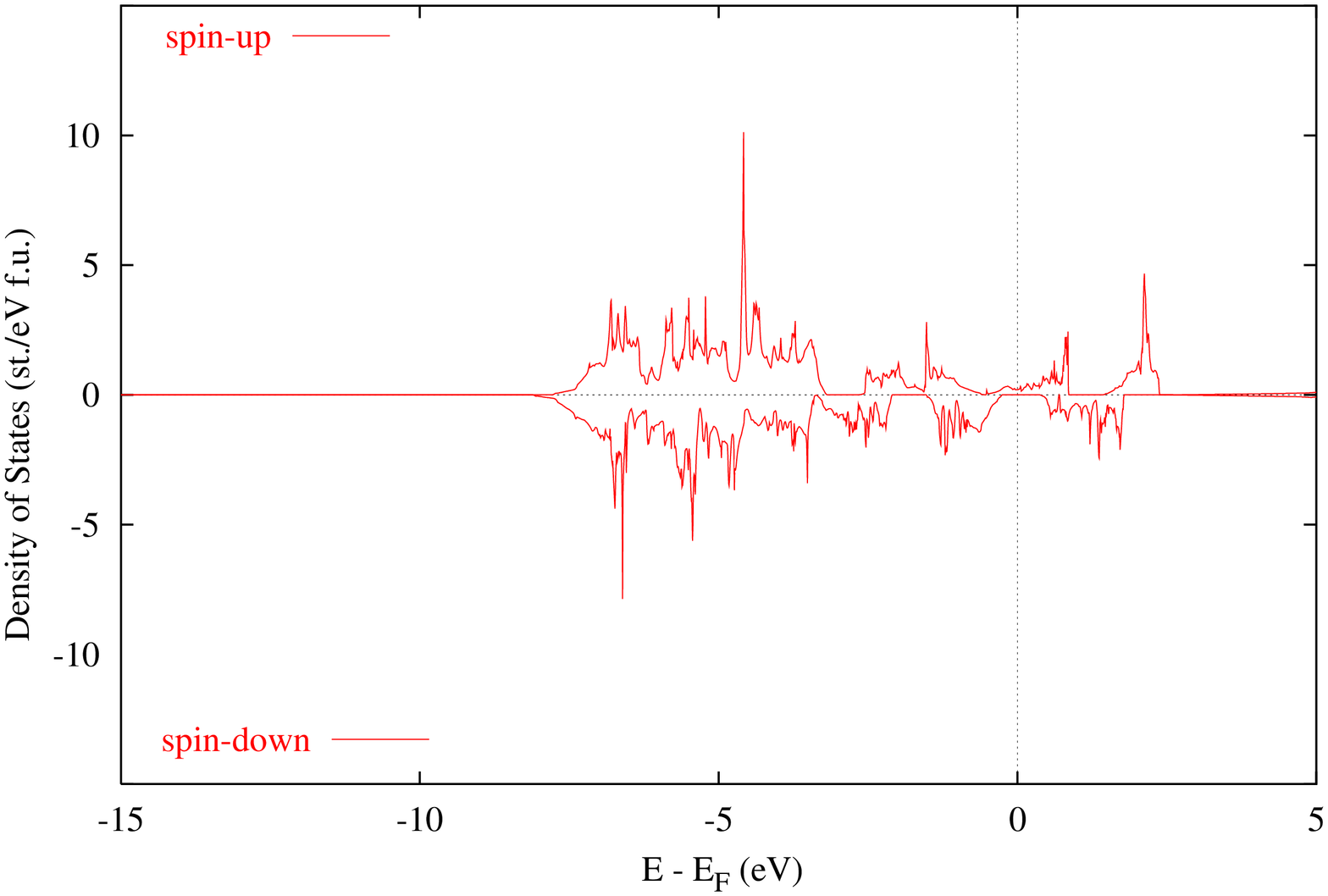}&
\includegraphics[scale=.37,angle=-90]{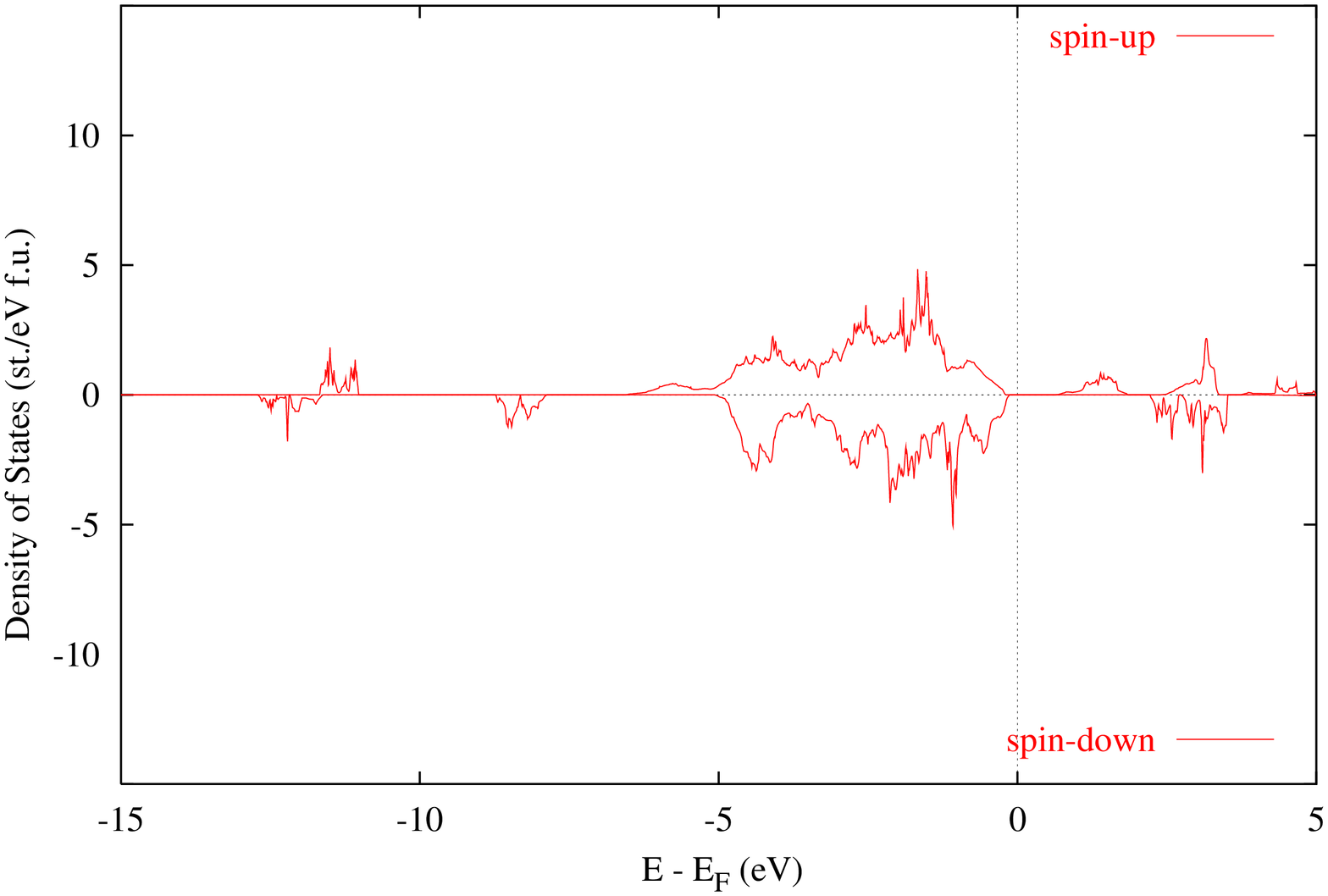}
\end{tabular}
\caption{Spin-polarized total (top row), Fe-(middle row) and O-(bottom row) 
partial densities of states
from the LSD calculation (left column) in comparison with the SIC-LSD (right column) counterparts 
for the Verwey charge ordered phase in the inverse spinel structure.
The partial DOS for empty spheres are of no significance and therefore have
not been shown.}
\label{Fig1}
\end{figure*}

\begin{figure*}
\begin{tabular}{cc}
\includegraphics[scale=.37,angle=-90]{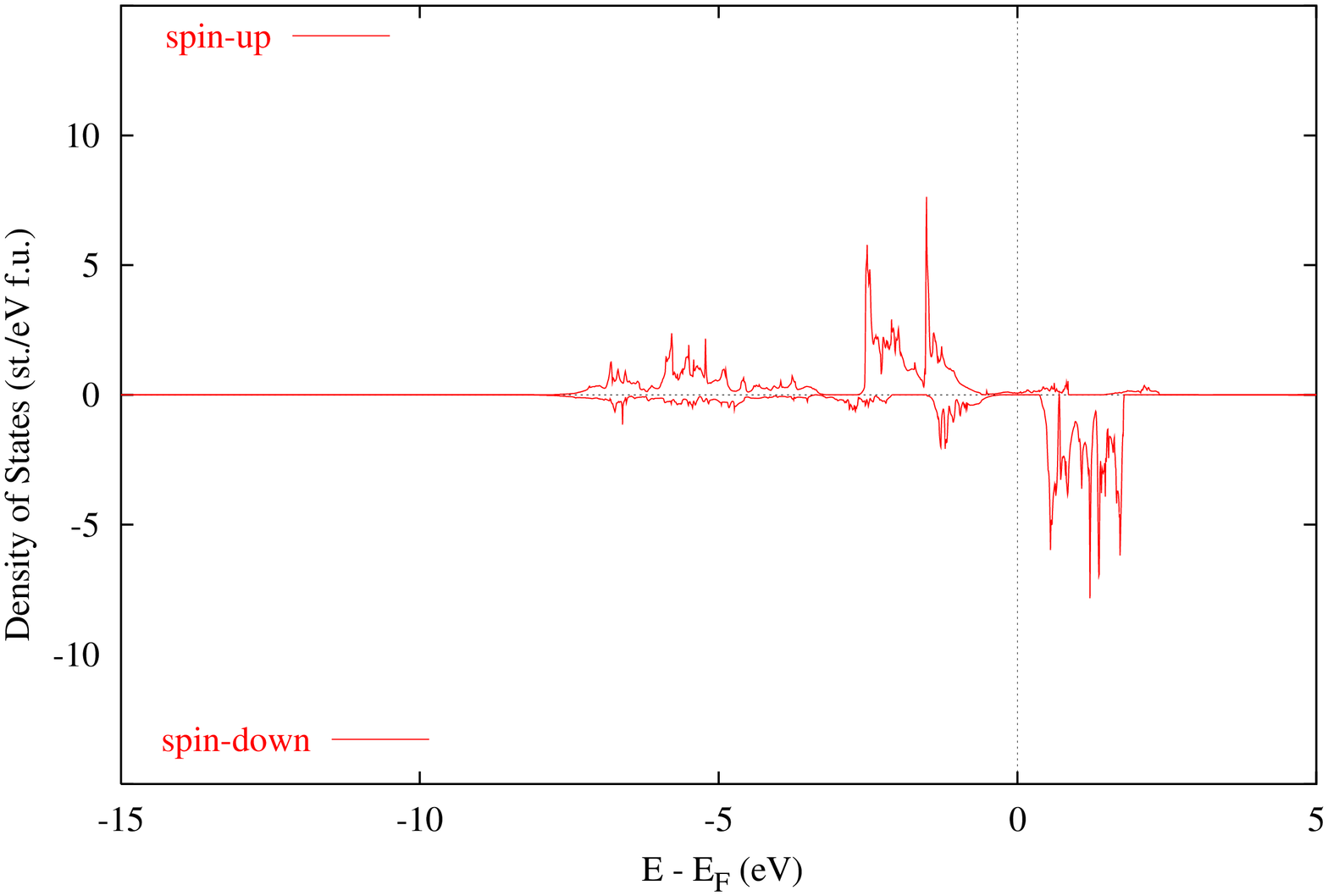}&
\includegraphics[scale=.37,angle=-90]{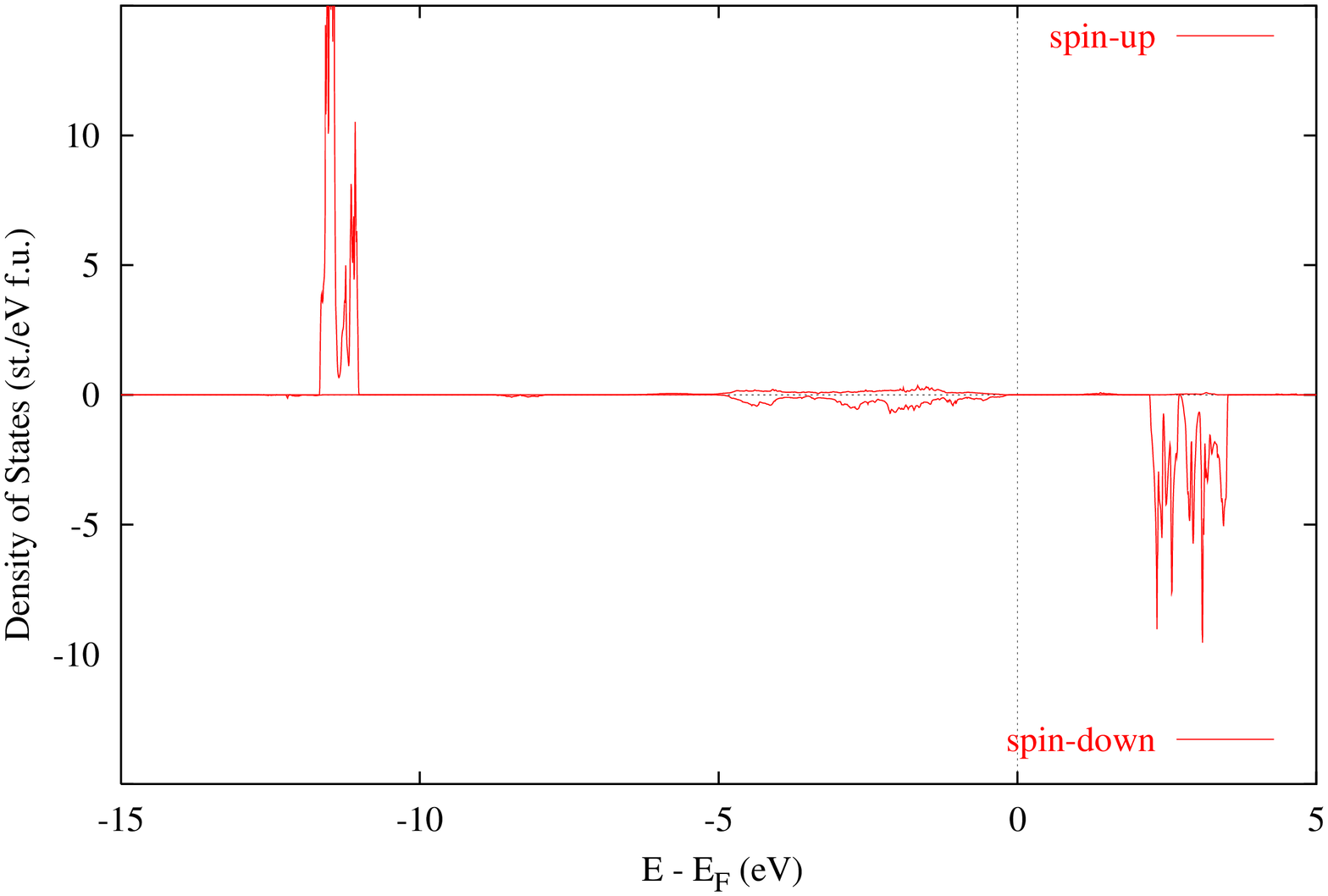}\\
\includegraphics[scale=.37,angle=-90]{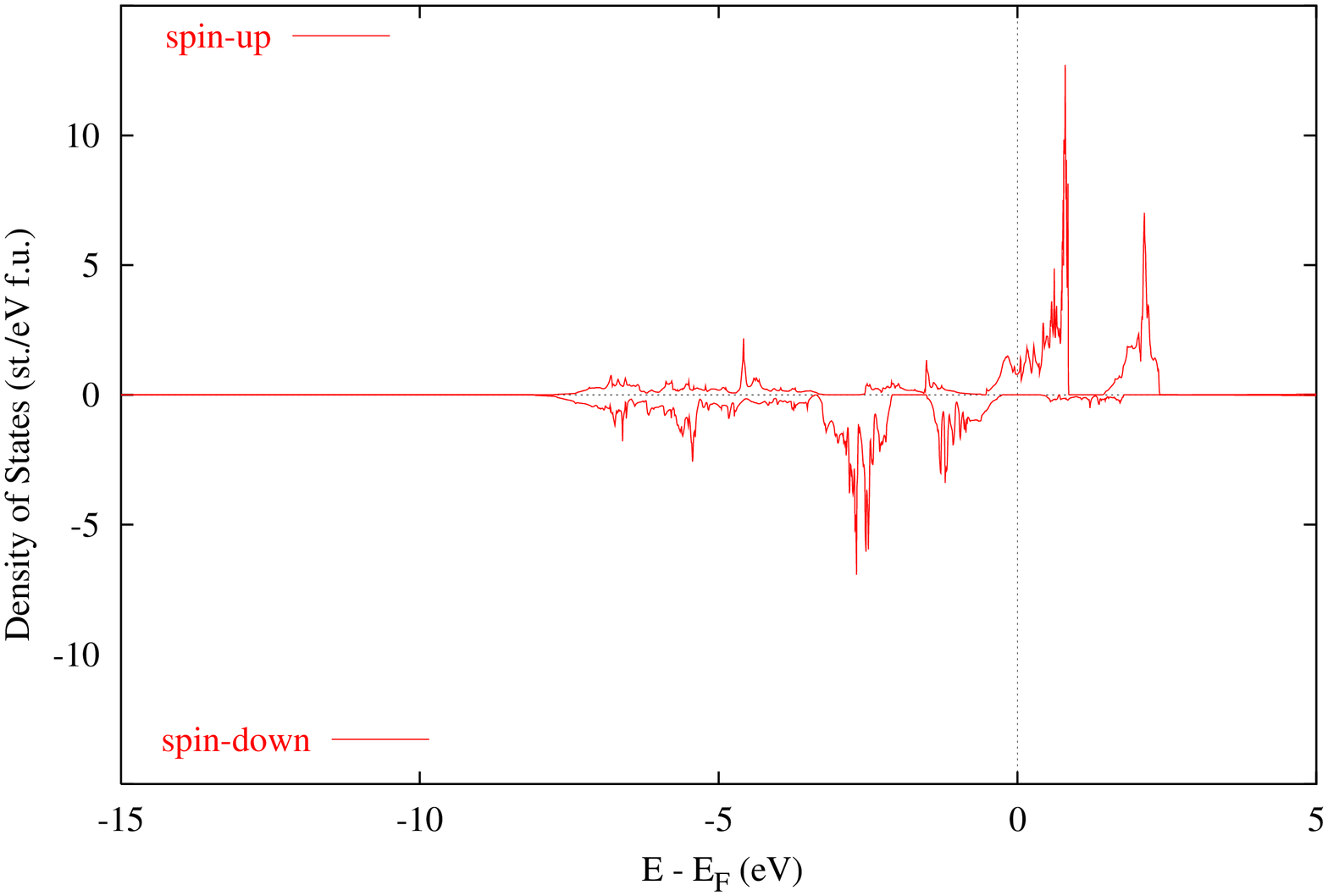}&
\includegraphics[scale=.37,angle=-90]{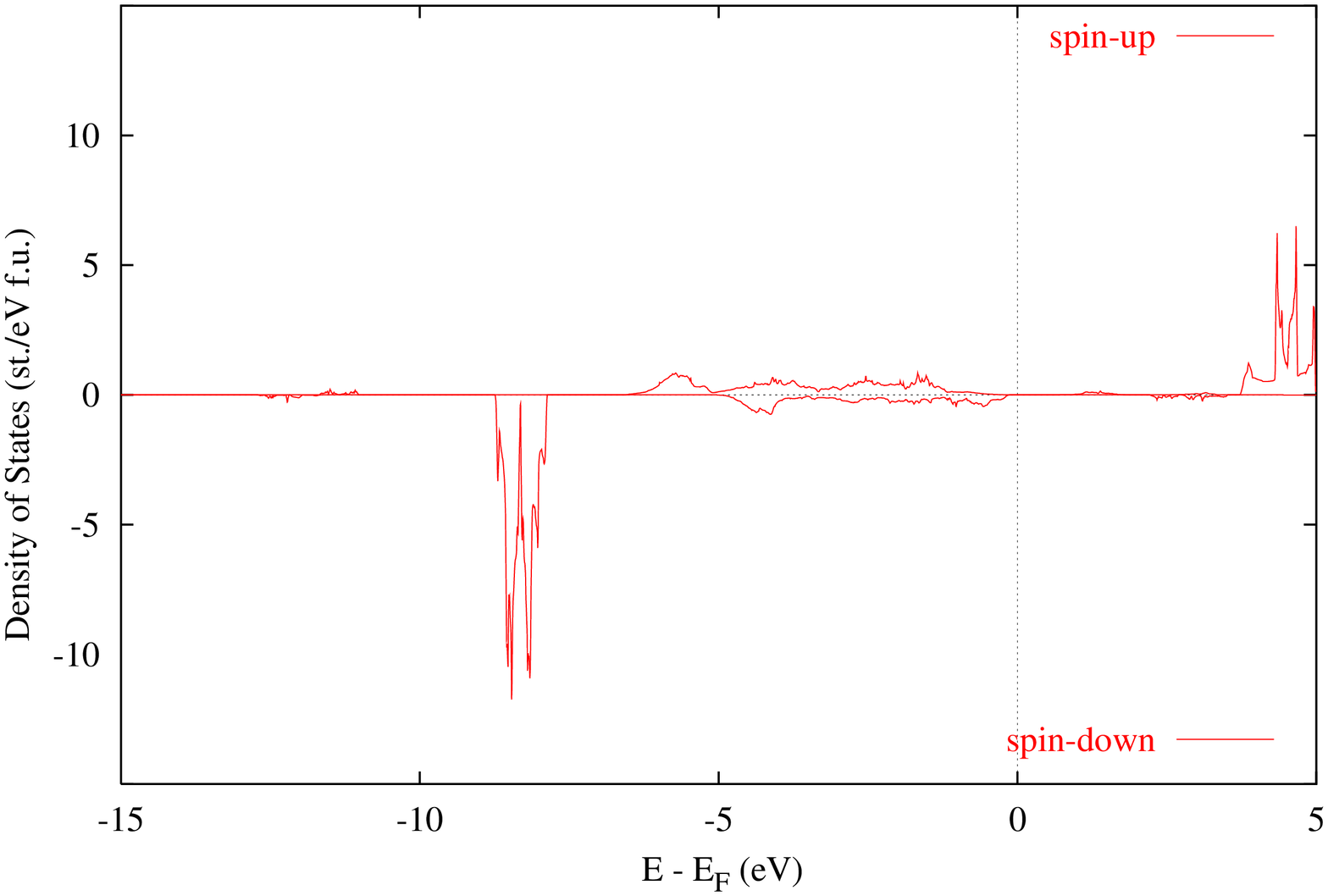}\\
&
\includegraphics[scale=.37,angle=-90]{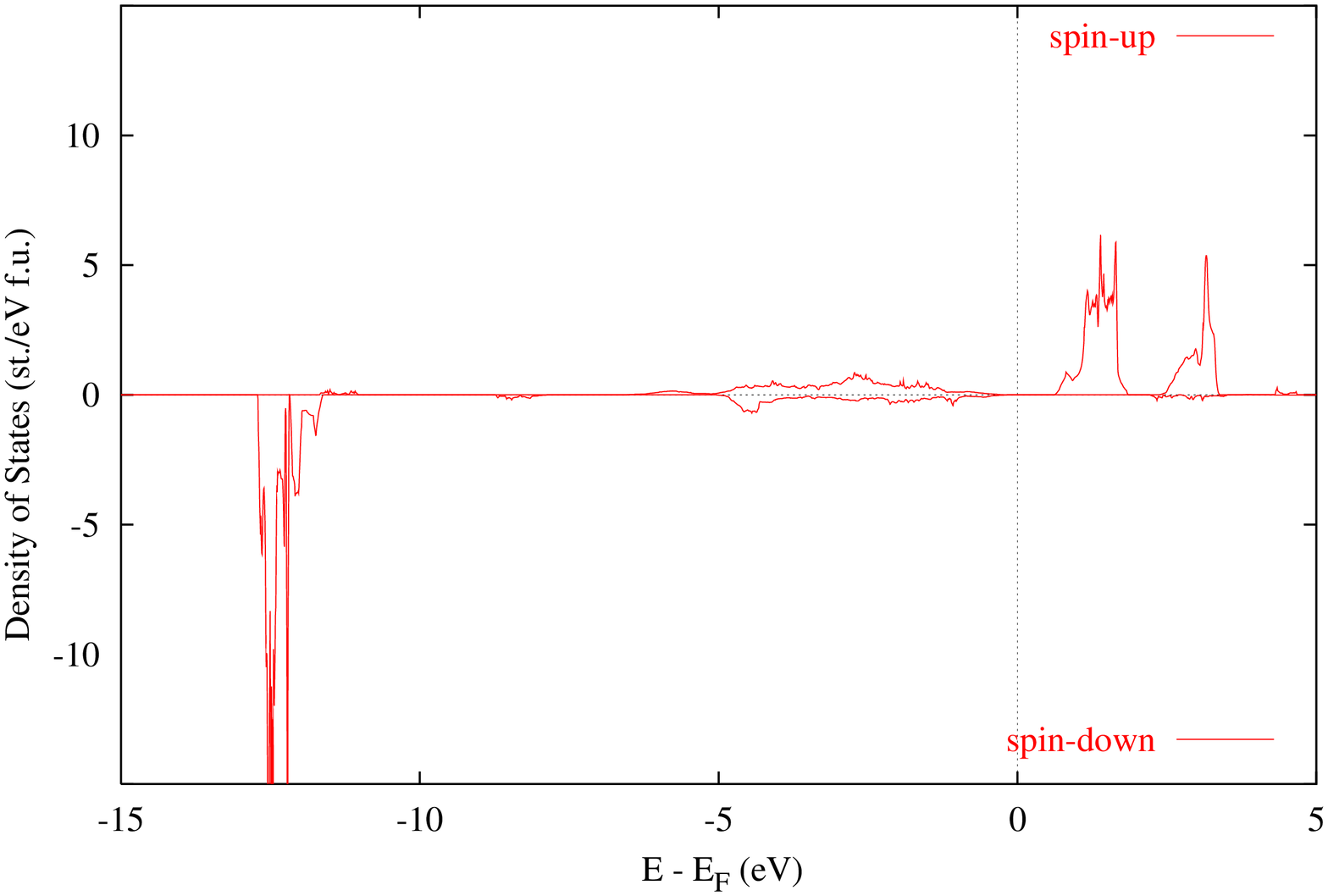}
\end{tabular}
\caption{Spin-polarized densities of states for the A-(top row) and B1(middle row) --Fe 
sites in the LSD calculation (left column) (note that in the LSD there is 
only one type of octahedral sites (B1$\equiv$B2$\equiv$B))
in comparison to the respective A-(top row), 
B1-(middle row) and B2-(bottom row)
Fe sites from the SIC-LSD calculation (right column) in the Verwey charge 
ordered phase.}
\label{Fig2}
\end{figure*}

\begin{figure}
\begin{tabular}{c}
\includegraphics[scale=.37,angle=-90]{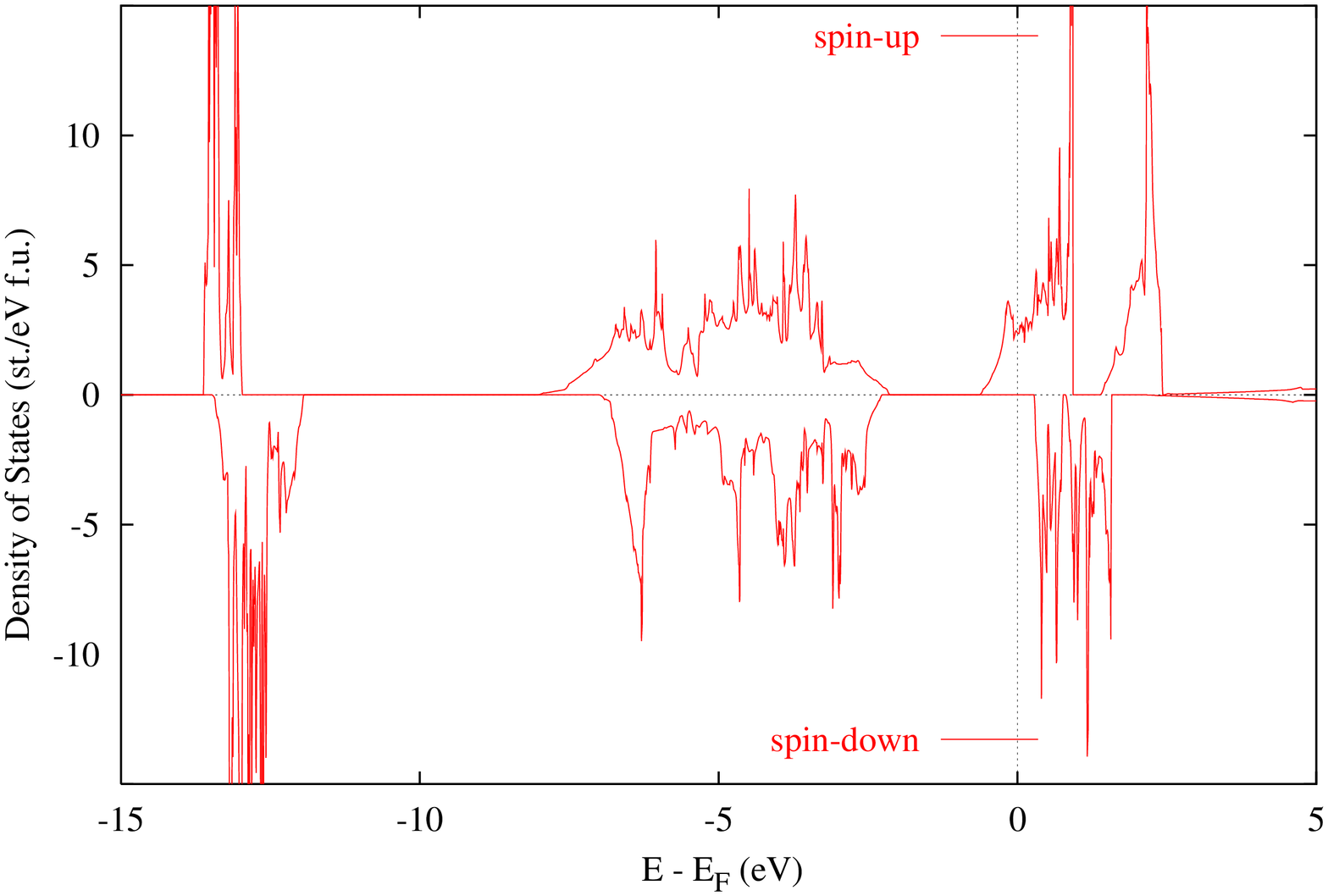}\\
\includegraphics[scale=.37,angle=-90]{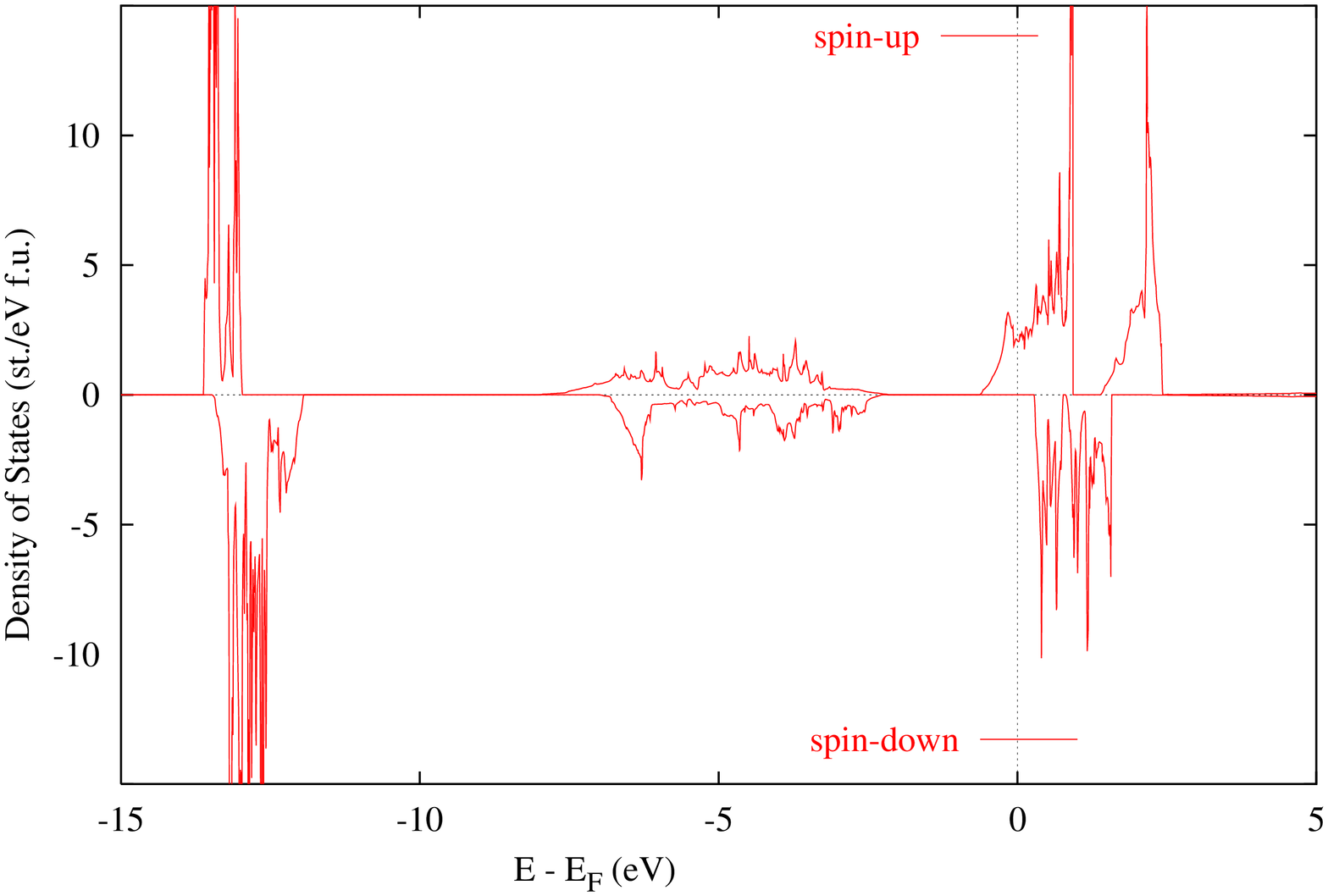}\\
\includegraphics[scale=.37,angle=-90]{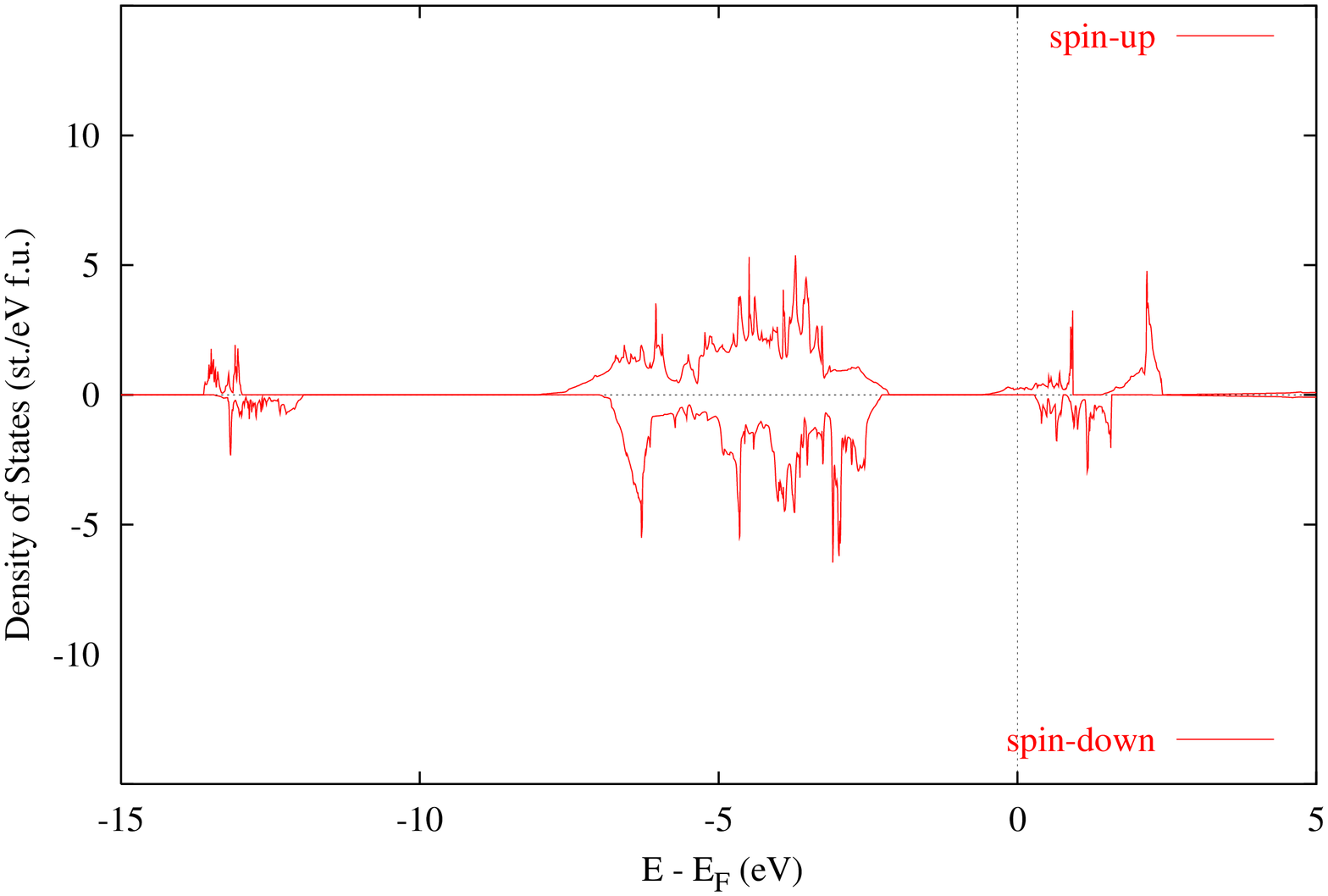}
\end{tabular}
\caption{ Spin-polarized total (top), Fe- (middle), and O-partial (bottom) densities 
of states for the second charge order scenario, namely with trivalent 
configuration on all Fe-sites in the SIC-LSD implementation.}
\label{Fig3}
\end{figure}

\begin{figure}
\begin{tabular}{c}
\includegraphics[scale=.37,angle=-90]{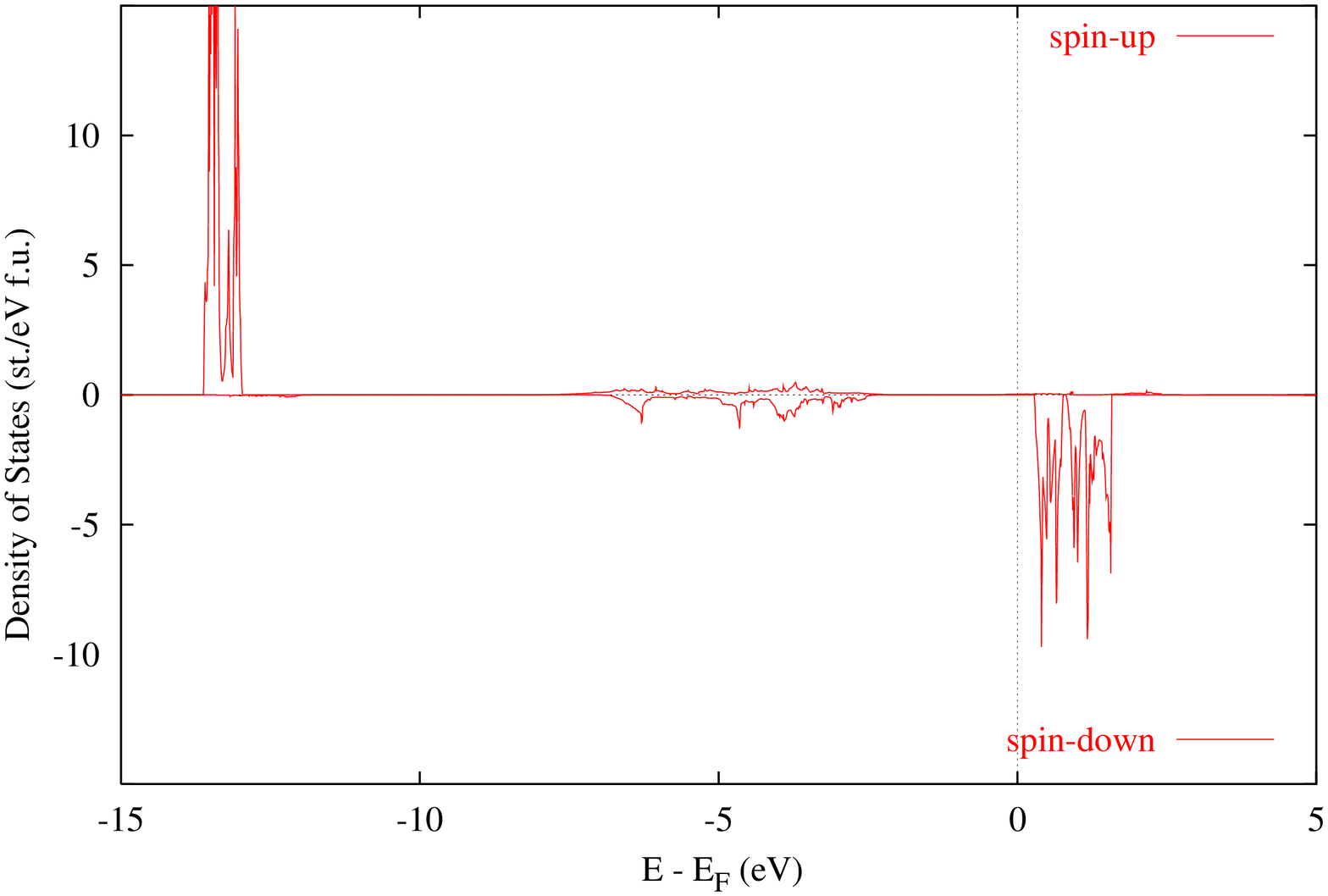}\\
\includegraphics[scale=.37,angle=-90]{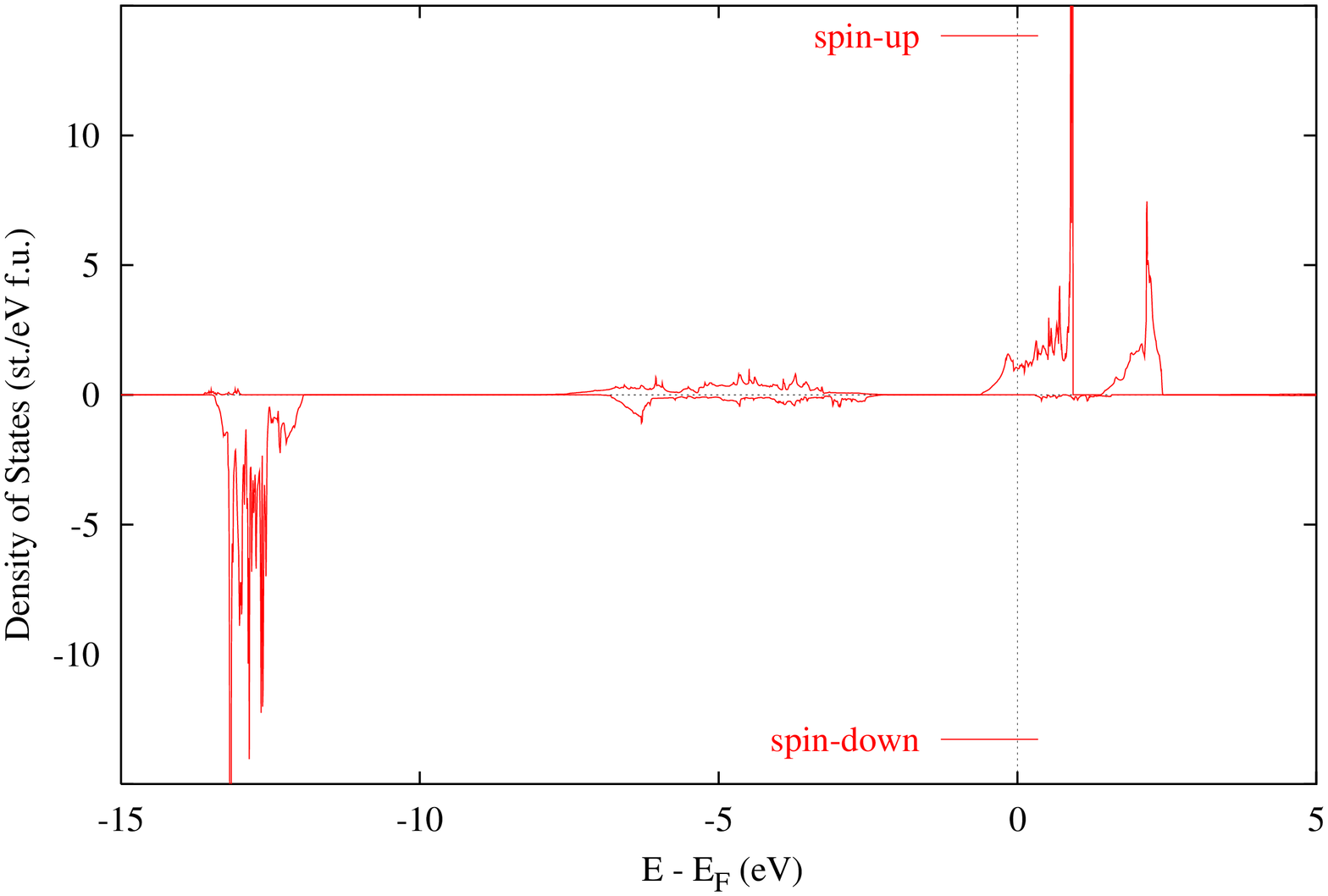}
\end{tabular}
\caption{Spin-polarized densities of states for the A- (top) and B1- (bottom)
Fe sites for the second scenario, namely with trivalent 
configuration on all, tetrahedral and octahedral, Fe-sites in the SIC-LSD 
implementation. Note that in this scenario 
there is only one type of octahedral sites (B1$\equiv$B2$\equiv$B).}
\label{Fig4}
\end{figure}

\end{document}